# Ageing under oscillatory stress: Role of energy barrier distribution in thixotropic materials §

Asheesh Shukla and Yogesh M. Joshi*

Department of Chemical Engineering, Indian Institute of Technology Kanpur,

Kanpur 208016, INDIA.

* Corresponding Author, E-Mail: joshi@iitk.ac.in

Telephone: +91 512 259 7993, Fax: +91 512 259 0104

§ Dedicated to Prof. Morton M Denn on the occasion of his 70th Birthday

## Abstract

In this work the ageing dynamics of soft solids of aqueous suspension of laponite has been investigated under the oscillatory stress field. We observed that at small stresses elastic and viscous moduli showed a steady rise with the elastic modulus increasing at a faster rate than the viscous modulus. However at higher stresses both the moduli underwent a sudden rise by several orders of magnitude with the onset of rise getting shifted to a higher age for a larger shear stress. We believe that the observed behavior is due to interaction of barrier height distribution of the potential energy wells in which the particle is trapped and strain induced potential energy enhancement of the particles. Strain induced in the material causes yielding of the particles that are trapped in the shallower wells. Those trapped in the deeper wells continue to age enhancing the cage diffusion timescale and thereby the viscosity which lowers the magnitude of strain allowing more particles to age. This coupled dependence of strain, viscosity and ageing causes forward feedback for a given magnitude of stress leading to sudden rise in both the moduli. Changing the microstructure of the laponite suspension by adding salt affected the barrier heights distribution that showed a profound influence on the ageing behavior. Interestingly, this study suggests a possibility that any apparently yielded material with negligible elastic modulus, may get jammed at a very large waiting time.

## I. Introduction

Understanding structure and flow behavior of highly viscous pasty materials has significant importance in the chemical industry. These materials include concentrated suspensions (Courtland and Weeks 2003; Treece and Oberhauser 2007), emulsions (Hebraud et al. 1997; Mason et al. 1995), foams (Cantat and Pitois 2005), inks (Buron et al. 2004), waxy derivatives of the petroleum industry (Petersson et al. 2008), industrial slurries (Cristiani et al. 2005), drilling fluids (Lahalih and Dairanieh 1989), etc. The characteristic property that is common in all these materials is their time dependent viscosity leading to the thixotropic behavior (Barnes 1997). In addition, the microstructure of these materials evolves with time, a feature generally known as ageing and gets influenced by the deformation field. This brings a strong history dependence and hence leads to significant difficulties while studying the physical behavior of these materials. In this work we will investigate effect of oscillatory stress field on the ageing behavior and study how the microstructure affects the same.

The above mentioned soft materials that show thixotropic behavior are generally represented as soft glassy materials (Bandyopadhyay et al. 2006; Cipelletti and Ramos 2005; Coussot 2006; Coussot 2007; Fielding et al. 2000; Sollich et al. 1997). In these materials due to crowding of constitutive entities, the translational diffusive motions of the same are strongly restricted. This renders a limited access to the phase space and the system falls out of equilibrium. Non-ergodic character of the system induces ageing wherein the arrested entities undergo the microscopic dynamics within the cage and lower the potential energy (Wales 2003). For molecular glasses the process of ageing, followed by a rapid decrease in the temperature below the glass transition temperature, leads to progressive ordering in the system, which increases the density of the same (McKenna 2003; Struik 1978). However, in the colloidal glasses, since the system is ergodic at the molecular length-scale and nonergodic only at or above the particle length-scale, increase in the bulk density does not take place. The ageing dynamics continues until a thermodynamic equilibrium or a saturated glassy state is reached wherein no further ageing takes place (O'Connell and McKenna 1999).

The ageing dynamics in the soft glassy materials is strongly dependent on the microstructure of the non-ergodic state. Every arrested particle in the jammed system has energetic interactions with its surrounding constituents that confine it in a cage-like environment. This state of an arrested particle can be represented by an energy well. An application of the deformation field increases the potential energy of a trapped particle and facilitates its diffusion out of the trap. Usually there exists a distribution of the energy barriers associated with the traps and hence all the arrested entities do not get uniformly affected by the deformation field (Di Leonardo et al. 2005; Fielding et al. 2000; Ianni et al. 2007; Viasnoff and Lequeux 2002).

It is generally observed that the deformation field retards the ageing dynamics. However, in order to observe the effect of deformation field on the ageing process, the timescale of the deformation field needs to be faster than the relaxation time of the material (Di Leonardo et al. 2005). If applied stress is above the yield stress, ageing stops; while for the intermediate stresses the dependence of the dominating timescale of the material on age weakens (Abou et al. 2003; Cloitre et al. 2000; Coussot et al. 2006; Derec et al. 2000; Derec et al. 2003; Joshi and Reddy 2008). In this work we study effect of stress controlled oscillatory shear on the ageing behavior of aqueous suspension of laponite, a model soft glassy material. Most of the previous studies have employed the continuous deformation field (Abou et al. 2003; Di Leonardo et al. 2005) or the continuous stress field (Cloitre et al. 2000; Joshi and Reddy 2008; Joshi et al. 2008; Ovarlez and Coussot 2007) to analyze the ageing behavior under shear. The benefit of using oscillatory stress field is two fold. Since the imposed flow field is stress controlled, deformation induced in the material is determined by the viscoelastic character of the same at that instant. Furthermore, since the flow field is oscillatory; it disturbs the system over a limited range of the deformation. Thus, in a stress controlled oscillatory field the induced oscillatory deformation corrects itself as ageing progresses. We show that probing the ageing dynamics by oscillatory stress field gives useful insights into the interaction between barrier heights associated with the energy wells in which the particles are trapped, corresponding cage diffusion time scales and the flow field.

**II. Material, sample preparation and viscometry**

Laponite is a synthetic hectorite clay and belongs to the structural family known as the 2:1 phyllosilicates (Van Olphen 1977). Laponite is an important additive used in the chemical and the food industry to control rheological behavior of the end product.[1] Laponite is composed of disc shaped particles with a diameter 25 nm and a layer thickness 1 nm (Kroon et al. 1998). The chemical formula for laponite is $Na_{+0.7}[(Si_8Mg_{5.5}Li_{0.3})O_{20}(OH)_4]_{-0.7}$. In a powder form laponite particles are present in the stacks of layers with sodium atoms in the interlayer gallery. In an aqueous medium, dissociation of the sodium ions renders a net negative charge to its surface. The edge of a particle is composed of hydrous oxide and is less negative in the basic pH medium while positive in the acidic pH medium (Tawari et al. 2001; Van Olphen 1977). At pH 10 the negative charge on the surface leads to overall repulsion among the laponite particles (Bonn et al. 1999; Joshi et al. 2008; Tanaka et al. 2005; Tanaka et al. 2004), although the attractive interactions between the edge and the surface can not be ruled out (Mongondry et al. 2005). Soon after mixing laponite with water, system enters into a non ergodic state (Joshi 2007). Addition of NaCl increases concentration of cations that screen the negative charge on the surface. Consequently, increase in the concentration of salt causes enhanced attraction among the laponite particles. Aqueous suspension of laponite shows a rich phase behavior and various versions of its phase diagrams with respect to concentration of laponite and that of salt have been proposed in the literature (Michot et al. 2006; Mongondry et al. 2005; Mourchid et al. 1995; Ruzicka et al. 2006; Tanaka et al. 2004).

Laponite RD used in this study was procured from Southern Clay Products, Inc. The white powder of Laponite was dried for 4 hours at 120 °C before mixing it with water at pH 10 having predetermined concentration of NaCl. The basic pH of

---

[1] http://www.laponite.com

10 was maintained by the addition of NaOH to provide chemical stability to the suspension (Mourchid and Levitz 1998). The suspension was stirred vigorously for 15 min and left undisturbed for 2 weeks in a sealed polypropylene bottle. The schematic of the experimental procedure is described in figure 1. In this work, we have carried out oscillatory shear experiments using a stress controlled rheometer, AR 1000 (Couette geometry, bob diameter 28 mm with gap 1mm). For an each test, shear cell was filled up with the independent sample and subjected to an oscillatory deformation with very high stress amplitude to carry out shear melting. The suspension yielded under such a high stress and eventually showed a plateau of low viscosity that did not change with time. Shear melting was necessary to achieve a uniform initial state. We stopped the shear melting experiment at this point in time, from which the aging time was measured. Subsequent to shear melting, we carried out the oscillatory shear experiments by applying different shear stress amplitudes in the range 0.01 Pa to 40 Pa with frequency 0.1 Hz to record the ageing behavior of this system under shear. In this work we have used 2.8 weight % (≈ 1.1 volume %) suspension of laponite at two concentrations of NaCl leading to $C_s$ =0.1 mM and $C_s$ =5mM. $C_s$ is the molar concentration of $Na^+$ ions of the aqueous medium prior to the addition of laponite. Former concentration of $C_s$ =0.1 mM is due to addition of NaOH to maintain pH of 10. To avoid the loss of water by evaporation or the possibility of $CO_2$ contamination of the sample, the free surface of the suspension was covered with a thin layer of low viscosity silicon oil during the course of viscometric measurements. All the experiments reported here were carried out 20 °C.

### III. Results

The experimental protocol involved shear melting before carrying out the oscillatory test in order to reach the uniform initial state in every experiment. In the shear melting experiment, under application of very high shear stress amplitude, the sample yielded and eventually resultant strain amplitude reached a plateau. In figure 2 we have plotted strain amplitude for two systems as a function of time. In the shear melting experiments, it is generally observed that once a plateau of high strain amplitude (low viscosity) is reached, which does not change with time, the ageing behavior of the system becomes independent of the minor differences and the time elapsed during the shear melting experiment (Cloitre et al. 2000; Joshi et al. 2008).

Subsequent to shear melting, experiments we applied oscillatory stress field of varying magnitude to the sample. Figure 3 shows evolution of the viscous ($G''$) and the elastic ($G'$) modulus with age for different shear stress amplitudes for a system without salt. At low stress amplitudes, both the moduli gradually increased with respect to waiting time (age). It can be seen that $G'$ increased at a faster rate and eventually crossed $G''$. As the stress amplitude increased, evolution of $G''$ with waiting time showed progressively sluggish increase at small waiting times while progressively rapid increase later. On the other hand, at higher stresses, value of $G'$ was below the detection limit of the rheometer indicating sample to be in a liquid state. At a certain waiting time $G'$ showed a sharp increase of several orders of magnitude over a short span of waiting time on a double logarithmic scale. We represent this waiting time as a critical waiting time, which increased with increase in the stress amplitude. We have plotted the corresponding primary harmonic in strain as a function of waiting time in figure 4. It can be seen that the primary harmonic in strain was higher for the larger magnitude of stress amplitude. Figure 4 also describes harmonic analysis of the oscillatory data at 40 Pa stress amplitude, wherein the ratio of the third harmonic in strain to the primary harmonic in strain is plotted as a function of ageing time. It can be seen that the third harmonic in strain was always less than 20 % of the primary harmonic in strain. Apparently similar behavior is also observed for variety of other soft materials (Krishnamoorti and Giannelis 2001; Wyss et al. 2007) and appears to be a generic characteristic of these class of systems. This behavior justified the representation of the age dependent viscoelastic behavior in terms of $G'$ and $G''$ in these large amplitude oscillatory shear experiments. Figure 3 illustrates that increase in the stress amplitude not only deferred the critical waiting time at which sharp rise in $G'$ was observed to higher values, but also enhanced the sharpness of the increase on a double logarithmic scale. We have quantified this sharpness by plotting $d\ln G'/d\ln t_w$ at $G'=1$ Pa, as a function of magnitude of oscillatory stress, as an inset, in figure 5. It can be seen that, in the limit of small shear stress (limiting case of linear response), $d\ln G'/d\ln t_w$ reaches a plateau, which can be considered to be a reference value. This means that, in this limit, evolution of elastic modulus does not depend on the magnitude of oscillatory stress. For higher values of stress, however, $d\ln G'/d\ln t_w$ can be seen to be increasing vary rapidly. Figure 5 also shows that the time taken for $G'$ to cross over $G''$ (as represented by $t^*$) increased with $\sigma_0$. The corresponding elastic modulus at the cross over ($G'=G''$) showed an exponential increase with $\sigma_0$. Interestingly at 40 Pa stress amplitude, both the moduli $G'$ and $G''$ increased very sharply almost parallel to each other. In figure 5, we have also plotted the waiting time at which elastic modulus became measurable (≈1 mPa), which showed a significant rise with a power law exponent of 5 with respect to the shear stress amplitude.

Figure 6 reports ageing under various shear stress amplitudes for 2.8 wt. % laponite suspension having 5 mM NaCl. The nature of evolution of both elastic and viscous moduli with waiting time was very different than that of a system without salt as shown in figure 3. It can be seen that up to stress amplitude of 20 Pa, the qualitative nature of evolution of $G'$ and $G''$ was very similar, with $G'$ being significantly larger than $G''$. However at 30 Pa stress, the value of $G'$ was negligibly small to be detected by the rheometer for up to waiting time of 26000 s (~7 h). We have plotted corresponding strain amplitude (primary harmonic) as a function of waiting time in figure 7. We have also plotted ratio of third harmonic in strain to primary harmonic in strain as a function waiting time that shows the third harmonic in strain was always less than 20 % of the primary harmonic in strain. Similar to the behavior of the system having no salt, application of higher magnitude of stress increased the magnitude of strain; however its effect on elastic and viscous modulus was very different from the system without salt, as demonstrated in figure 3 and 6.

**IV. Discussion**

The aqueous suspension of laponite, particularly above concentration of 1 volume %, is known to undergo ergodicity breaking soon after preparation. Such behavior is a result of enhanced volume (effective volume) of the particle by a factor of 60 due to electrostatic screening length (Joshi 2007). Addition of salt reduces the electrostatic screening length and enhances the attraction among the particles, which causes the nonergodic state of the system to become less homogeneous with increase in the concentration of salt (Ruzicka et al. 2008). In the literature, some groups represent the state of the system having no salt as a repulsive glass (Schosseler et al. 2006; Tanaka et al. 2004) while some state it as a homogeneous gel (Mongondry et al. 2005). However, irrespective of the state in which the system is arrested, the particle (arrested entity) can be considered to be trapped in a potential energy well created by its interactions with the surrounding particles. In this state, the particles undergo the microscopic dynamics favoring those structural rearrangements that take them to a lower potential energy state (Wales 2003). By virtue of the location, the orientation of the particle and the nature of cage, energy associated with each of the particle is different. Therefore there exists a natural distribution of potential energies and barrier heights associated with the wells in which the particles are trapped (Bouchaud 1992; Sollich 1998). This distribution is expected to be a function of concentration of laponite and that of salt, age of the sample and deformation history of the sample. The timescale over which the particle diffuses out of the energy well is known as cage diffusion time scale (or $\alpha$ relaxation mode). Thus, by virtue of distribution of barrier heights, there also exists a distribution of $\alpha$ relaxation modes. Fielding et al. (Fielding et al. 2000) proposed that an application of deformation with magnitude $l$ increases the potential energy of the particle by the magnitude $kl^2/2$. If this value exceeds the barrier height (or yield energy), particle overcomes the barrier and a yielding event occurs.

Subsequent to the yielding event, particle gets trapped into a new cage and starts the ageing dynamics afresh. If deformation is continuous and sufficient to cause continuous yielding, particle does not undergo ageing. Thus, if oscillatory deformation is of magnitude $l$, then all the particles trapped in a well with barrier height less than $kl^2/2$ will undergo continuous yielding. However all those particles that are trapped in the wells with barrier heights greater than $kl^2/2$ will not yield and continue to age thereby reducing their potential energy.

In the present experimental protocol, we are controlling the amplitude of the oscillatory stress. Therefore the viscoelastic nature of the system decides the magnitude of strain amplitude. In the shear melting step very high magnitude of shear strain gets induced in the material as shown in figure 2, which is expected to cause complete yielding of the sample. After the shear melting step, according to Fielding et al. (Fielding et al. 2000), freshly yielded elements or particles get trapped in energy wells with a certain distribution of energy barriers. We believe that this distribution is a function of physicochemical nature of the system, location and orientation of the particles, age and nature of deformation field. In a subsequent step, when shear stress of lower magnitude is applied to the system leading to significantly lower strain compared to that associated with shear melting, only those particles get arrested in the energy wells that have depths greater then the energy associated with the reduced strain they are experiencing. This scenario is schematically shown in figure 8(a). The gray line in the same figure represents natural distribution of energy wells in which particles will get trapped subsequent to shear melting step, while the distribution of particles trapped in the deeper wells (than ½ $kl^2$) is shown on the right hand side by a think black line. The remaining particles that are undergoing yielding (having energy $O(k_B T)$) are represented by a thick vertical line on left hand side. The fraction of arrested particles is now free to undergo ageing dynamics and thereby lower its energy as it is not yielded by the induced strain. The decrease in energy will cause enhancement in the cage diffusion time scale of the un-yielded fraction of the particles which will in turn bring about a gradual increase in the overall viscosity. The increase in viscosity will progressively lower the strain induced in the material as observed in figure 4. With a gradual decrease in the magnitude of strain, the number of particles that are trapped in deeper wells will also gradually increase and their ageing will cause increase in the rate at which viscosity increases. Eventually this coupled relation between strain, viscosity and ageing will cause a positive feedback leading to a sudden decrease in strain. This autocatalytic dynamics is expected to bring about a sharp increase in both elastic and viscous moduli as observed in figure 2. We also observe that the sudden increase in storage modulus occurred at lower waiting time (or age) with decrease in the amplitude of shear stress. Figure 4 shows lowering of initial strain with decrease in applied stress, which causes a larger number of particles to get arrested in the deeper well as shown in figure 8 (a). This accelerates the overall ageing dynamics bringing about the sudden rise in storage modulus at earlier waiting time (age). Therefore above discussion also points towards an important possibility that any apparently shear rejuvenated (completely yielded) material with negligible elastic modulus, which does not show any variation of viscous properties over a prolonged period, may get jammed at a very large waiting time. Interestingly Coussot *et al.* (Coussot et al. 2002) also observed sudden rise in viscosity in creep flow with critical time at the onset of rise increasing with increasing applied shear stress. They develop a qualitative model which relates the observed behavior to the dependence of viscosity to organization and disorganization of the network caused by ageing and rejuvenation. However in the present study, we owe this behavior to distribution of barrier heights and its interaction with the deformation field.

Unlike the demonstration of sharp increase in elastic and viscous moduli in the system without any salt, the system with 5 mM NaCl demonstrated a gradual increase up to stress amplitude of 20 Pa as shown in figure 6. For the stress amplitude of 30 Pa, the sample showed a liquid-like behavior up to an explored waiting time of 26000 s (for no salt system, for 30 Pa stress $G'$ showed a sharp rise at waiting time of 11000 s). The possible explanation for this behavior is represented by a schematic shown in figure 8(b) in which distribution of energy well depths is plotted. We observed that with increase in shear stress amplitude from 20 Pa to 30 Pa (with corresponding increase in initial strain as shown in figure 7), the ageing behavior of laponite suspension having 5 mM salt was altogether different. Such behavior can arise from narrow distribution of the energy well depths (gray line), shown in the figure 8(b), such that energy corresponding to strain induced for applied oscillatory stress of 20 Pa and 30 Pa encompass the major portion of the distribution dome. This observation suggests that when transition from shear melting step to the ageing step occurs, energy associated with initial strain induced for the oscillatory stress of 20 Pa is low enough to let significant fraction of the particles to get trapped in the deeper well. However, for 30 Pa of oscillatory stress, induced strain raises energy of the particles thereby most of the particles are rejuvenated as shown in figure 8(b).

As discussed before, addition of NaCl induces attraction among the particles. It is observed that increase in the salt concentration caused increase in the correlation length (Nicolai and Cocard 2001). Recent small angle x-ray scattering study has also suggested more homogeneous phase for the system having no salt compared to the one at the higher concentration of salt (Ruzicka et al. 2008). By virtue of lower correlation length (no salt system), an individual arrested particle has many nearest neighbors than the suspension having 5 mM salt. Furthermore, compared to a low ionic concentration state of aqueous suspension of laponite, the interactions among the particles at high concentration of salt of the same are primarily attractive in origin. Since the nearest neighbor interactions are generally the most dominating, and given that for a 5 mM system, those are fewer and mainly attractive in origin, distribution of barrier heights for the trapped particles can be expected to be narrower than for a no salt system which has more nearest neighbors and attractive as well as repulsive interactions. This observation suggests that the suspension having 5 mM salt should get affected over a narrow range of strain amplitude (and hence stress magnitude) compared to the suspension with no salt as indeed observed in the present experiments. Interestingly our recent creep-recovery experiments showed less ultimate recovery for systems with higher salt concentrations (Reddy and Joshi 2008) suggesting narrower relaxation time distribution with increase in the concentration of salt (Joshi 2009).

In general, we observed that the microstructure affects the ageing dynamics under shear through interaction between the potential energy enhanced by induced strain and the distribution of energy barriers in which the particles are trapped. Consequently, the microstructure of the system gets affected as deformation field causes yielding of the particles trapped in the shallower wells letting the particles trapped in the deeper wells age. In the industrial soft glassy materials this coupled relationship between structure and deformation field gets more complicated due to far more complex microstructure of the real systems. Understanding how the industrial soft materials behave under the deformation field is very important and certainly more experimental and theoretical studies are required for the effective processing and mixing of these materials at the industrial scale.

## V. Conclusions

In this work we investigated the ageing behavior of a model soft glassy material, an aqueous suspension of laponite, under oscillatory stress field. We observed that when no salt is present in the system, ageing under shear was accompanied by a sharp increase in both elastic ($G'$) and viscous ($G''$) moduli. An increase in the shear stress amplitude not only deferred the increase in $G'$ and $G''$ to a higher value of age but also enhanced the sharpness of the increase on a double logarithmic scale. We believe that this behavior is due to the presence of distribution in the barrier heights of the potential energy wells in which the particles are trapped. In a stress dependent oscillatory deformation field, maximum strain induced in the material is determined by its viscoelastic character. Such induced strain enhances the potential energy of the trapped particles. If this enhancement is sufficient to overcome the energy barrier, the particles yield. Therefore a particular deformation field causes yielding of only a part of the system letting the remaining part to age. Partial ageing of the system causes continuing increase in the viscosity, which gradually decreases the strain amplitude that allows a larger portion of the system to age, increasing the rate at which viscosity increases. This process auto-catalytically leads to a sharp increase in elastic and viscous moduli. Therefore, this study suggests a possibility that any apparently yielded material that has negligible elastic modulus and does not show any variation in viscous properties over a prolonged period, may get jammed at a very large waiting time. For a system with 5 mM NaCl, the increase in both the moduli was more gradual and the system was affected by a narrower range of the stress amplitudes than that of the system with no salt. The observed behavior may arise because of the narrow distribution of barrier heights of the potential energy wells caused by the microstructural changes brought about in the system by the addition of salt. This study also suggests a possibility that any apparently yielded material that has negligible elastic modulus and does not show any variation in viscous properties over a prolonged period, may get jammed at a very large waiting time.

**Acknowledgement:** This work was supported by BRNS young scientist research project awarded by Department of Atomic Energy, Government of India to YMJ.

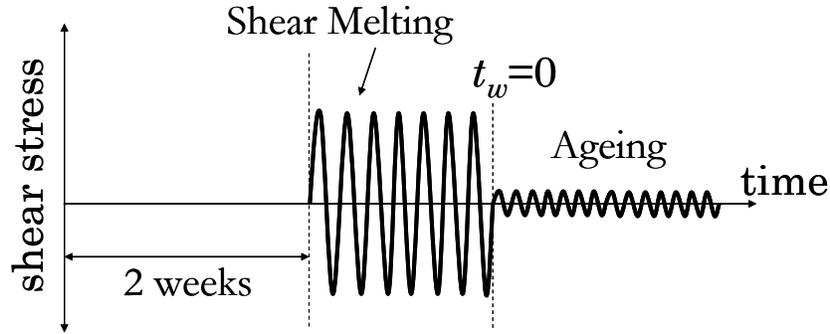

**Figure 1.** Schematic of an experimental procedure. After keeping the sample idle for two weeks complete rejuvenation was carried out by applying high shear stress amplitude in the shear melting step. Subsequently shear stress amplitude of varying magnitude was applied in the ageing step to study ageing behavior under oscillatory shear stress.

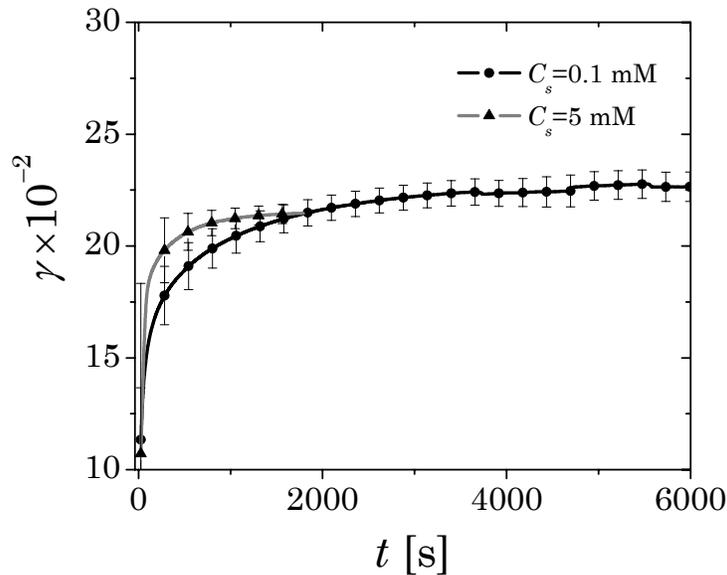

**Figure 2.** Variation of strain amplitude as a function of time in the shear melting step. Shear stress amplitude of 70 Pa with frequency 0.1 Hz was applied to system with $C_s$ =0.1 mM (filled circles), while stress amplitude of 65 Pa with frequency 0.1 Hz was applied to system with $C_s$ =5 mM (filled tringles).

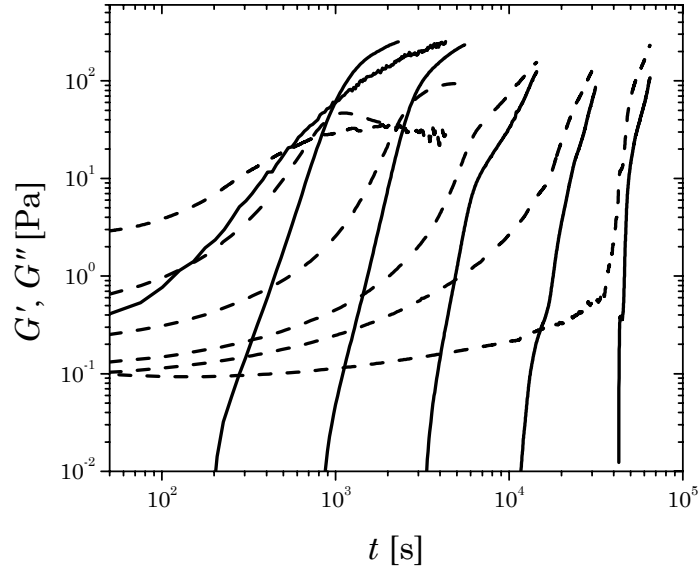

**Figure 3.** Evolution of elastic ($G'$) and viscous ($G''$) modulus as a function of magnitude of the applied shear stress for 2.8 weight % aqueous laponite suspension having no salt ($C_s$ =0.1 mM). From left to right shear stresses: 0.01 Pa, 15 Pa, 20 Pa, 25 Pa, 30 Pa and 40 Pa. Dashed lines represent the viscous modulus while the solid lines represent the elastic modulus.

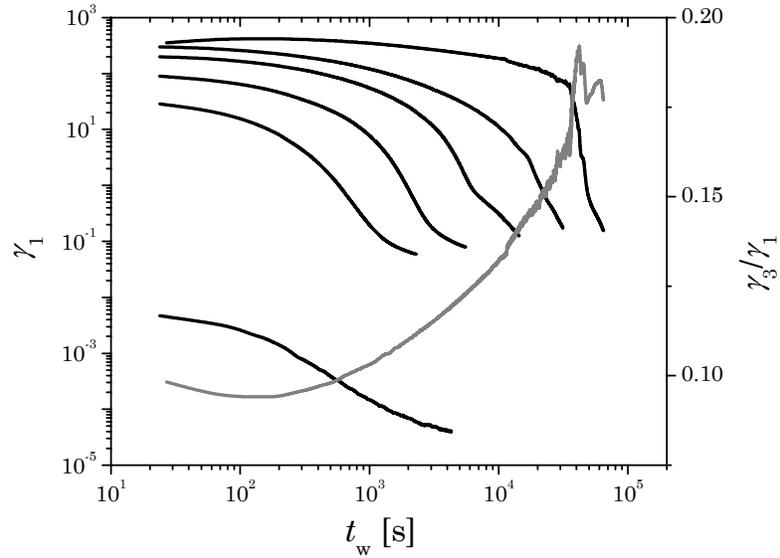

**Figure 4.** Evolution of primary harmonic in strain (black lines) plotted as a function of age for various stresses for 2.8 weight % laponite suspension without salt. From top to bottom, shear stresses: 40 Pa, 30 Pa, 25 Pa, 20 Pa, 15 Pa, and 0.01 Pa. Ratio of third harmonic in strain to primary harmonic in strain (thick gray line) is also plotted against the ageing time for stress amplitude of 40 Pa. It can be seen that the magnitude of third harmonic in strain is at least 20 % of the magnitude of primary harmonic in strain.

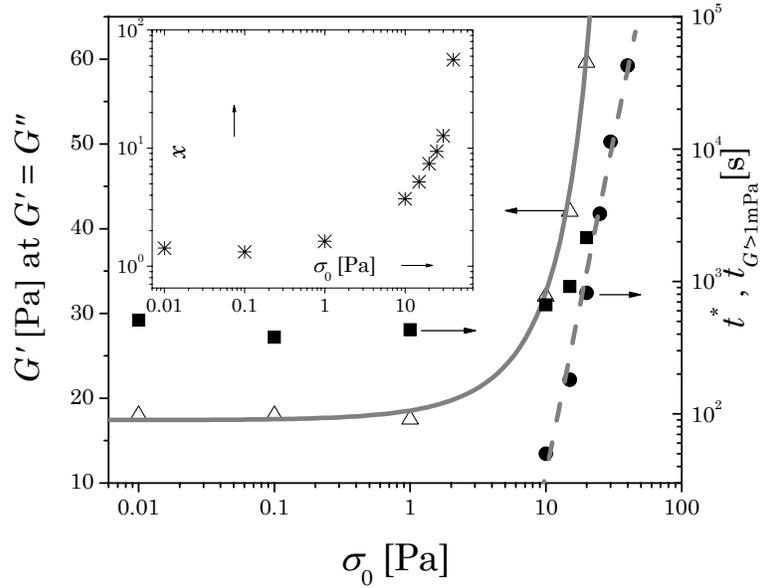

**Figure 5.** The value of elastic modulus ($G'$) and age ($t^*$) at $G'=G''$ is plotted against magnitude of the oscillatory shear stress for 2.8 weight % sample without salt. Open triangles represent the modulus while the filled squares represent $t^*$. Filled circles represent the age at which elastic modulus becomes measurable (≈ 1mPa). Gray line represents an exponential fit to the modulus-stress relationship, while dashed gray line represents power law fit given by $t_{G'>1mPa} \sim \sigma_0^5$. Inset shows $d\ln G'/d\ln t_w \,(=x)$ at $G'=1$ Pa, as a function of magnitude of oscillatory stress.

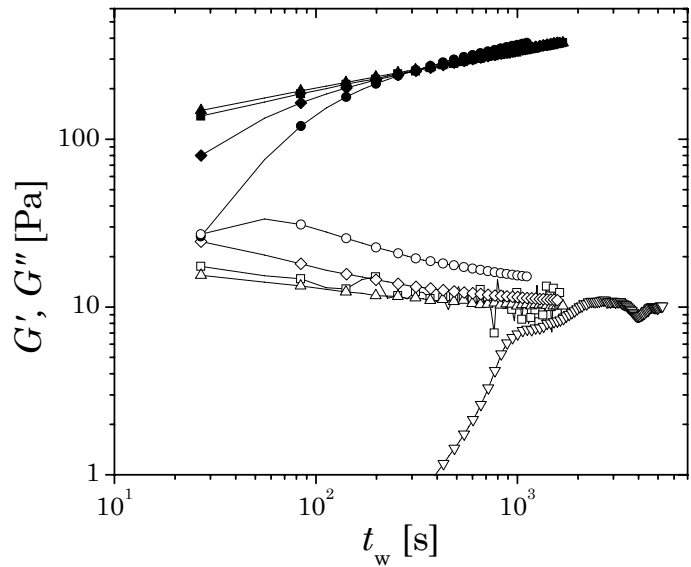

**Figure 6.** Evolution of elastic ($G'$) and viscous ($G''$) modulus with age for various magnitudes of the applied shear stress for 2.8 weight % aqueous laponite suspension with $C_s$=5 mM (down triangles 30 Pa, circles 20 Pa, diamonds 10 Pa, up triangles 1 Pa and squares 0.1 Pa). Open symbols represent the viscous modulus while the filled symbols represent the elastic modulus. For the stress amplitude of 30 Pa, $G'$ was negligibly small to be detected by the rheometer throughout the explored age up to 26000 s (~7 hr, complete range is not shown in the figure).

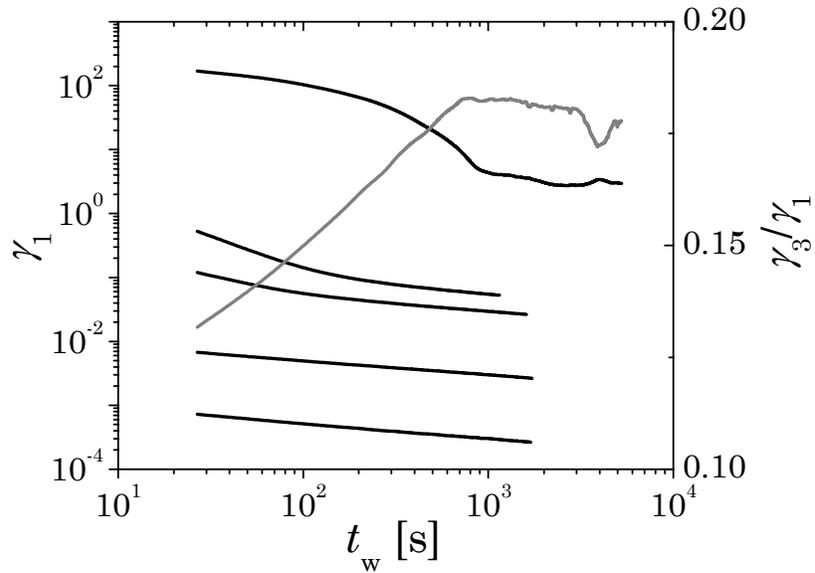

**Figure 7.** Evolution of primary harmonic in strain (black lines) plotted as a function of age for various stresses for 2.8 weight % laponite suspension with 5 mM salt. From top to bottom, shear stresses: 30 Pa, 20 Pa, 10 Pa, 1 Pa, and 0.1 Pa. Ratio of third harmonic in strain to primary harmonic in strain (thick gray line) is also plotted against the ageing time for stress amplitude of 30 Pa. It can be seen that the magnitude of third harmonic in strain is at least 20 % of the magnitude of primary harmonic in strain.

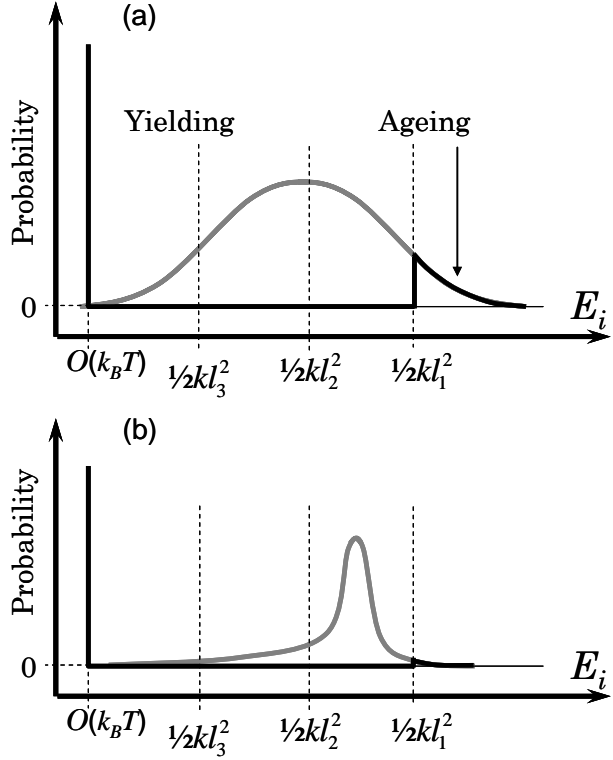

**Figure 8.** Schematic of distribution of energy barriers after sample undergoes a transition from shear melting to ageing (figure 1). The ordinate represents the probability that any randomly chosen particle is trapped in a well having depth (barrier height) $E_i$. The gray line represents natural distribution of energy wells in which particles will get trapped squbsequent to the shear melting step. In shear melting step the particles are expected to have energy of the order of $k_B T$. In the subsequent ageing experiments, lower magnitude of strain causes fraction of particles, by virtue of their surrounding environment, to get arrested in the energy wells having depth greater than energy associated with the corresponding strain, $\frac{1}{2} k l^2$ (represented by a distribution shown in dark thick line towards right hand side). The remaining particles undergo continuous yielding and have energy $O(k_B T)$ (represented by the vertical line on the left hand side). Figures (a) and (b) represent possible scenarios with interaction of energy associated with strains of various magnitudes and broad and narrow distributions of energy well depths respectively.